\documentclass[twocolumn,showpacs,preprintnumbers,superscriptaddress,amsmath,amssymb]{revtex4}
\begin{document}
\title{Information-carrying Hawking radiation and the number of microstate for a black hole}
\author{Qing-yu Cai}
\email{qycai@wipm.ac.cn}
\affiliation{State Key Laboratory of Magnetic Resonances and Atomic and Molecular Physics,
Wuhan Institute of Physics and Mathematics, Chinese Academy of Sciences, Wuhan
430071, China}
\author{Chang-pu Sun}
\email{cpsun@csrc.ac.cn}
\affiliation{Beijing Computational Science Research Center, Beijing, 100084, China}
\affiliation{Collaborative Innovation Center of Quantum Information and Quantum Physics,
University of Science and Technology of China, Hefei, Anhui 230026, China}
\author{Li You}
\email{lyou@mail.tsinghua.edu.cn}
\affiliation{State Key Laboratory of Low Dimensional Quantum Physics,
Department of Physics, Tsinghua University, Beijing 100084, China}
\affiliation{Collaborative Innovation Center of Quantum Matter, Beijing, China}

\begin{abstract}

We present a necessary and sufficient condition to falsify whether a Hawking
radiation spectrum indicates unitary emission process or not from the perspective of information theory.
With this condition, we show the precise values of Bekenstein-Hawking
entropies for Schwarzschild black
holes and Reissner-Nordstr\"om black holes can be calculated by counting the microstates
of their Hawking radiations. In particular, for the extremal Reissner-Nordstr\"om black hole,
its number of microstate and the corresponding entropy we obtain are found to
be consistent with the string theory results.
Our finding helps to refute the dispute about the Bekenstein-Hawking entropy
of extremal black holes in the semiclassical limit.
\end{abstract}


\maketitle

\section{Introduction}

In the earlier 1970s, black hole thermodynamics were established
based on the analogies between the laws of black hole dynamics
and thermodynamics~\cite{dc70,pf71,sh71,jb73,bch73,swh74,swh75}.
With this arises two open questions, the information loss paradox
and the puzzled origin for a black hole's entropy.
When quantum theory is applied to a black hole,
 Hawking discovered a black hole radiates and the so-called Hawking
radiation is approximately thermal \cite{swh74,swh75}.
Since thermal radiations do not carry correlations,
the discovery of Hawking radiation leads to the famous claim that information about the collapsed
matter in a hole is lost~\cite{swh76,jp92}.
Recently, we present a plausible yet consistent resolution
for the black hole information loss paradox, in a series of studies~\cite{zcyz09,zczy111,zcy11,zczy132}
with collaborators.
At the heart of our work is the discovery of correlations among
Hawking radiations \cite{iy10,vac11,cc13,cc14} when the emission spectra
takes the nonthermal form of Parikh and Wilczek~\cite{pw00}.

The concept of entropy is not only used in information theory,
it is a relevant concept also in thermodynamics, where the precise value
for the entropy of a system is determined by the number of its microstates~\cite{etj57}.
The existence of information default, or an amount of information
unreachable to an observer, is the origin for a black hole's entropy.
The nonzero entropy for a black hole implies the lack of information about
the collapsed matter in a hole for an observer outside the hole.
A long-standing conjecture relates the entropy of a black hole to its microstates
near event horizon~\cite{jwy83,zt85,gth90,stu93}.
Bekenstein estimated it for a Schwarzschild black hole by considering
the change of its horizon surface area from a falling particle across the horizon
forty years ago~\cite{jb73}.
Based on information theory, he arrived at the
area entropy of a black hole $(\ln2/8\pi)kc^3A/G\hbar$,
where $A$ is the event horizon area of the black hole.
Bekenstein's result is almost the same as the standard result
${kA}/{4l_p^2}$ given by Hawking, which is obtained based on the thermodynamic relationship
between energy, temperature, and entropy~\cite{swh75}.
This establishes the entropy of a black hole is proportional to
its area at event horizon. The origin of this entropy, however, remains a puzzle until now.

Finding the precise value of the entropy for a black hole from counting its number
of microstates near the event horizon represents a daunting task~\cite{iw08}.
Accomplishing this goal will shine light on the puzzled origin for a black hole's entropy
and implicate the establishment of a successful quantum gravity theory.
Serious efforts have been carried out along this direction without too much successes.
Most notably, it was found that the entropies for some black holes,
such as the five-dimensional extremal black holes can be obtained
in string theory by counting the degeneracies of the BPS soliton bound
states~\cite{sv96}. An approach based on the correspondence
between ${\rm AdS_3}$ and conformal field theory was also proposed to
count the number of states for the BTZ black hole~\cite{as98}.
Other studies have extended the above calculations to entropies for
extremal or near-extremal black holes~\cite{sl05}. With limited
exciting progresses made for some types of black holes as mentioned above,
the relationship between the number of
microstates and the Bekenstein-Hawking entropy remains to be established
for non-extremal black holes, including the most commonly discussed Schwarzschild black hole.

This work is focused on counting the microstates of Hawking radiation
to recover the Bekenstein-Hawking entropy for a black hole in the semiclassical limit.
We first discuss a necessary and sufficient condition capable of falsifying
unitary Hawking radiation from the perspective of information theory.
The precise values of Bekenstein-Hawking entropies for Schwarzschild black holes and
Reissner-Nordstr\"om black holes are then calculated by counting the number
of microstates of their respective Hawking radiations.
For Reissner-Nordstr\"om black holes, alternative spectra for neutral
or charged Hawking radiation emissions are considered, and based on which the
number of microstates for the extremal Reissner-Nordstr\"om black hole is counted
and its nontrivial entropy obtained.
Finally, we speculate that the number of microstates for a variety of extremal
black holes can be probed by analyzing their alternative Hawking radiation spectra
for the corresponding black holes in the semiclassical limit.
For simplicity, units with $\hbar=c=k=G=1$ are adopted in the following.

\section{information-carrying Hawking radiation}
\label{black:hole:entropy}

In information theory, entropy measures the amount of uncertainty for a variable.
It also measures the amount of information we gain when the variable becomes known.
The quantitative relationship between entropy and information
for a closed physical system can be expressed as $\Delta I=-\Delta S$,
where $\Delta I$ denotes information gained from a system
(through measurements) and $\Delta S$ denotes the corresponding decrease
for the amount of uncertainty of the closed system~\cite{lb56}. This presumption clearly
applies when the closed system dynamics are unitary as described by a Hamiltonian.

According to the celebrated no-hair theorem \cite{wi87},
the entropy $S(M)$ for a Schwarzschild black hole is a function
of its mass $M$ only. Its event horizon area $A=4\pi M^2$
is given by the mass as well. After a Hawking radiation
(a particle) $\omega_1$ is emitted from a black hole,
the entropy for the remaining black hole becomes $S(M-\omega_1)$.
The amount of information the first emitted particle $\omega_1$ carries off
is $I(\omega_1)=-[S(M-\omega_1)-S(M)]=S(M)-S(M-\omega_1)$.
The second Hawking radiation, a particle $\omega_2$,
subsequently emits following the first one. Afterwards the entropy for the black hole decreases
 to $S(M-\omega_1-\omega_2)$. Information the second particle
carries off, conditional on the first emission $\omega_1$ carries with it
the associated information $I(\omega_1)$, is easily computed to be
$I(\omega_2|\omega_1)=I(\omega_1,\omega_2)-
I(\omega_1)=-[S(M-\omega_1-\omega_2)-S(M-\omega_1)]=S(M-\omega_1)-S(M-\omega_1-\omega_2)$,
where $I(\omega_1,\omega_2)$ denotes the total amount of information carry away by the
two particles $\omega_1$ and $\omega_2$ (after $\omega_1$). This gives
$I(\omega_1,\omega_2)=I(\omega_1)+I(\omega_2|\omega_1)=S(M)-S(M-\omega_1-\omega_2)$.
On the other hand, if the same black hole (of mass $M$) were to emit
a single particle with energy $\omega_1+\omega_2$, the black hole's entropy
decreases to $S(M-\omega_1-\omega_2)$. The amount of information
this one particle of energy ($\omega_1+\omega_2$) carries out of the
hole is simply given by $I(\omega_1+\omega_2)=S(M)-S(M-\omega_1-\omega_2)$.
{\it For information to be conserved in Hawking radiation, or for Hawking radiation to be unitary},
from the above relationships,
we find
\begin{eqnarray}
I(\omega_1,\omega_2)=I(\omega_1)+I(\omega_2|\omega_1)=I(\omega_1+\omega_2),
\label{bdary}
\end{eqnarray}
is required.

Shannon entropy $S=-\sum_np_n\ln p_n$ measures the amount of
uncertainty for a closed system, where $p_n$ refers to the probability
for the $n$-the outcome. For a micro-canonical ensemble in equilibrium,
of which all microstates are equally likely, or $p_n\equiv p$,
its Shannon entropy reduces to Boltzmann entropy $S(p)=-\ln p$,
where $p=1/\Omega$, and $\Omega$ denotes the total number of microstates
for the finite system with a fixed total energy (or a closed system whose energy is conserved).
The black hole and its Hawking radiations together form a closed system,
which remains in equilibrium according to black hole thermodynamics.
Its entroy is therefore described by the Boltzmann entropy $S(p)=-\ln p$.
The condition (\ref{bdary}) above can be reexpressed as
\begin{eqnarray}\label{boundary:condition}
\Gamma(\omega_1,\omega_2)=\Gamma(\omega_1)\Gamma(\omega_2|\omega_1)=\Gamma(\omega_1+\omega_2),
\end{eqnarray}
where $\Gamma(\omega)$ is the probability for the emission of a Hawking radiation
particle $\omega$ from a black hole.
{\it The boundary condition in Eq.~(\ref{boundary:condition}) needs to be satisfied
if Hawking radiation is unitary since its derivation is based on
Eq.~(\ref{bdary}) which assumes conservation of information}.

The joint probability for all Hawking radiations (labeled with increasing integer indices)
from a black hole is simply
\begin{eqnarray}\label{joint:probability}
\Gamma(\omega_1,\omega_2,\cdots,\omega_n)=&&\Gamma(\omega_1)\cdot\Gamma(\omega_2|\omega_1)\notag
\cdots \\&&\Gamma(\omega_n|\omega_1,\omega_2,\cdots,\omega_{n-1}).
\end{eqnarray}
Iteratively using the boundary condition Eq. (\ref{boundary:condition}), we find an elegant formula
\begin{eqnarray}\label{sum:probability}
\Gamma(\omega_1+\omega_2+\cdots+\omega_n)=\Gamma(\omega_1,\omega_2,\cdots,\omega_n).
\end{eqnarray}
Making use of the Boltzmann entropy formula $S(p)=-\ln p$,
we find
\begin{eqnarray}\label{entropy:formula}
S(M)&=&-\ln \Gamma(\omega_1+\omega_2+\cdots+\omega_n) \notag
\\&=&-\ln\Gamma(\omega_1,\omega_2,\cdots,\omega_n),
\end{eqnarray}
where $M=\sum_{i=1}^{n}\omega_i$.
This shows that the boundary condition Eq.~(\ref{boundary:condition}) is sufficient
to protect information conservation, i.e.,
the total entropy $-\ln (1/\Omega(\omega_1,\omega_2,\cdots,\omega_n))=-\ln\Gamma(\omega_1,\omega_2,\cdots,\omega_n)$
is conserved during Hawking radiation process,
where $\Gamma(\omega_1,\omega_2,\cdots,\omega_n)=1/\Omega(\omega_1,\omega_2,\cdots,\omega_n)$.

Our discussions above thus show that (i) in order to conserve the total entropy
in Hawking radiation, Eq. (\ref{bdary})
or $\Gamma(\omega_1,\omega_2)=\Gamma(\omega_1+\omega_2)$
is required from the information theory perspective; and (ii)
the validity of $\Gamma(\omega_1,\omega_2)=\Gamma(\omega_1+\omega_2)$ enforces,
on the other hand, entropy conservation in Hawking radiation,
i.e., the total entropy remains a constant at any intermediate step
with a fixed number of Hawking radiations accompanied by a evaporating black hole,
including the final state composed of only Hawking radiations after a black is exhausted,
and the initial state of a black hole before any radiation is emitted.
{\it Thus, $\Gamma(\omega_1,\omega_2)=\Gamma(\omega_1+\omega_2)$
constitutes a necessary and sufficient
condition for conservation of entropy in Hawking radiation}.
This proves that Hawking radiation is unitary \emph{if and only if} (iff)
$\Gamma(\omega_1,\omega_2)=\Gamma(\omega_1+\omega_2)$ is satisfied.
In an earlier work~\cite{zcyz09}, it was obtained that the initial total number of microstates
(Bekenstein-Hawking entropy) for a Schwarzschild black hole is the same as the total number of microstates
(entropy) specified by all Hawking radiations with the nonthermal spectrum by Parikh and Wilczek~\cite{pw00},
which indeed satisfies the condition Eq.~(\ref{boundary:condition}). On the other hand,
it was shown that
one can get the Parikh-Wilczek nonthermal spectrum without referring to black hole geometry when
the condition Eq.~(\ref{boundary:condition}) is enforced~\cite{dcls14}.
These two sides combined thus show that the nonthermal spectrum of Parikh and Wilczek
is both sufficient and necessary to maintain entropy conservation in Hawking radiation
process in the semiclassical limit.

Black hole can serve as an ideal testing ground for quantum gravity theories,
all of which require unitary evolution.
To maintain unitarity for Hawking radiation, an emission spectrum will have to
satisfy the iff relationship Eq.~(\ref{boundary:condition}). Otherwise,
unitarity breaks down.
One of the well-known spectra is given by Hawking,
the so-called thermal radiation spectrum with $\Gamma_T(\omega)=\exp(-8\pi M\omega)$,
which does no satisfy the iff condition $\Gamma(\omega_1,\omega_2)=\Gamma(\omega_1+\omega_2)$
(as shown step by step in detail in Ref.~\cite{czzy12}).
As a result, we have the long-standing information loss paradox.
In this case, as was shown by many authors \cite{zczy132,zw82,page83},
the total entropy for a black hole increases with emission of thermal radiations.
For Hawking radiation spectra of more complicated forms,
the iff condition we obtain can be easily checked to falsify
the correctness of the corresponding quantum gravity theories.
As a further application along this direction,
we show below the number of microstates of black holes can be computed
by using specific Hawking radiation spectra that satisfy the iff condition Eq.~(\ref{boundary:condition}).

\section{Schwarzschild black hole}
\label{schwarzshild:black:hole:entropy}

In this section, we consider the number of microstate for
a Schwarzschild black hole. We show it can be calculated with
the Hawking radiation spectrum obtained by Parikh and
Wilczek~\cite{pw00}, which satisfies the iff
condition of Eq.~(\ref{boundary:condition}) as was verified before.

In principle, the number of microstate $\Omega_{\rm initial}(M)$ for a black hole
can be computed directly from its microscopic quantum states.
Unfortunately, this approach cannot be adopted because the exact form
of the quantum state for a black hole needs a quantum gravity theory
yet to be established.
We will use a different approach instead which is based on a general rule
regarding unitary quantum state evolution. We recall that the
number of microstates for an excited atom can be obtained from the
number of microstates of the emitted photon
and the level degeneracy at the lower energy atomic state.
In quantum gravity theory, black holes are regarded as highly excited states.
Each Hawking radiation emission causes a black hole to jump into a less excited state.
In this sense, we can simply compute the entropy for a black hole by
counting the number of microstates for its emissions or Hawking radiations.
We follow each individual emission process of the
complete queue of Hawking radiations which
exhausts a black hole step by step. Finally, the spacetime
becomes Euclidian as the ground state.
After the first emission with energy $\omega_1$,
the number of microsates (initial black hole) before emission
$\Omega_{\rm initial}(M)$ changes to $\Omega_{\rm BH}(M-\omega_1)$ (remaining black hole).
If we assume the number of microstates for the emitted particle is
$\Omega_{\rm radiation}(\omega_1)$, we obtain for this case
\begin{eqnarray}
\Omega_{\rm initial}(M)=\Omega_{\rm radiation}(\omega_1)\cdot \Omega_{\rm BH}(M-\omega_1). \notag
\end{eqnarray}
Proceeding with the next emission of a particle $\omega_2$, we find similarly
\begin{eqnarray}
\Omega_{\rm BH}(M-\omega_1)=\Omega_{\rm radiation}(\omega_2)\cdot \Omega_{\rm BH}(M-\omega_1-\omega_2), \notag
\end{eqnarray}
which leads to the following relationship between the numbers of the initial and final microstates
for a black hole after two emissions,
\begin{eqnarray}
\Omega_{\rm initial}(M)&=&\Omega_{\rm radiation}(\omega_1)\cdot
\Omega_{\rm radiation}(\omega_2) \notag
\\ &&\cdot\Omega_{\rm BH}(M-\omega_1-\omega_2). \notag
\end{eqnarray}
Continue along the Hawking radiation queue until the black hole is exhausted by its $n$ radiations,
the above relation translates into
\begin{eqnarray}\label{total:number:microstates}
\Omega_{\rm initial}(M)=\prod_{i=1}^n\Omega_{\rm radiation}(\omega_i),
\end{eqnarray}
which shows the number of microstates for a black hole can be computed
from the number of microstates for its Hawking radiations. This is a
self-consistent outcome when Hawking radiation is unitary.

Given the radiation probability for the an emission $\omega$ being $\Gamma(\omega)$,
the number of microstates for this radiation is simply
\begin{eqnarray}\label{single:number:rate}
\Omega_{\rm radiation}(\omega)=\frac{1}{\Gamma(\omega)}.
\end{eqnarray}
In the scenario of Hawking radiation as tunneling, quantum fluctuations
cause abundant particle pairs created and annihilated near event horizon,
where gravitational potential is strongest.
Hawking radiation occurs when the positive energy particle from a
pair created inside the horizon escapes out of the horizon through tunneling.
With the semiclassical approximation~\cite{pw00}, the radiation probability is obtained as
\begin{eqnarray}
\Gamma(\omega)=e^{-2{\rm Im}(I)}, \notag
\end{eqnarray}
where the imaginary part ($\rm Im$) of the action $I$ for a positive energy particle ($+\omega$)
tunneling outside crossing the horizon is given by
\begin{eqnarray}
{\rm Im}(I)={\rm Im}\int_{r_{in}}^{r_{out}}p_rdr,  \notag
\end{eqnarray}
where the lower and the upper bounds of integration specify
the beginning and ending coordinates $r_{in}$ to $r_{out}$
for the particle. With the Painlev\'e line element,
\begin{displaymath}
ds^2=-\left(1-\frac{2M}{r}\right)dt^2+2\sqrt{\frac{2M}{r}}dtdr+dr^2+r^2d\Omega^2,
\end{displaymath}
it was shown by Parikh and
Wilczek~\cite{pw00} that when back reaction (i.e., energy
conservation) is enforced, one arrives at
\begin{eqnarray}
{\rm Im}(I)=+4\pi\omega\left(M-\frac{\omega}{2}\right).  \notag
\end{eqnarray}
The pair creation may happen outside the horizon where Hawking radiation
corresponds to a negative energy particle tunneling
into a black hole. Likewise, the imaginary part of the action for the negative energy
particle ($-\omega$) crossing into the horizon is found to be
\begin{eqnarray}
{\rm Im}(I)=+4\pi\omega\left(M-\frac{\omega}{2}\right).  \notag
\end{eqnarray}
When both contributions from positive and negative particles are
included, the probability for Hawking radiation as tunneling becomes
\begin{eqnarray}
\Gamma(\omega)=e^{-2{\rm Im}(I)}=e^{-8\pi\omega(M-\omega/2)},
\label{p1pw}
\end{eqnarray}
which was first discovered by Parikh and Wilczek~\cite{pw00}.
It is easy to verify that this spectrum satisfies the iff condition
Eq.~(\ref{boundary:condition}), therefore it can be applied
to calculate the number of microstates for a Schwarzschild black hole.

With the above result Eq. (\ref{p1pw}), the tunneling rate for the $i$th radiation
$\omega_i$ can be obtained as
\begin{eqnarray}\label{general:tunnel:rate}
\Gamma(\omega_i)=e^{-8\pi\omega_i\left(M-\sum\limits_{j=1}^{i-1}\omega_j-\omega_i/2\right)}.
\end{eqnarray}
Combing Eqs. (\ref{total:number:microstates}), (\ref{single:number:rate}),
(\ref{general:tunnel:rate}), and the iff condition Eq.~(\ref{boundary:condition}),
the number of microstates for a Schwarzschild black hole with mass $M$ is found to be
\begin{eqnarray}\label{total:number}
\Omega_{\rm initial}(M)=\prod_{i}^{n}\frac{1}{\Gamma(\omega_i)}=e^{4\pi M^2}, \notag
\end{eqnarray}
which corresponds to an entropy of
\begin{eqnarray}
S=\ln \Omega_{\rm initial}(M)=4\pi M^2,  \notag
\end{eqnarray}
exactly equals to the value of Bekenstein-Hawking entropy for
the Schwarzschild black hole derived by Hawking~\cite{swh75}.

\section{Reissner-Nordstr\"om black hole}

The line element for the Reissner-Nordstr\"om black hole in the Painlev\'e coordinates is
\begin{eqnarray}
ds^2 &=&-\left(1-\frac{2M}{r}+\frac{Q^2}{r^2}\right)dt^2+2\sqrt{\frac{2M}{r}-\frac{Q^2}{r^2}}dtdr \notag \\
&&+dr^2+r^2d\Omega^2, \notag
\end{eqnarray}
which reveals a spacetime that is stationary and nonsingular at horizon,
thus can be applied to study particle tunneling straightforwardly~\cite{pw00}.
If only neutral particles are emitted from a Reissner-Nordstr\"om black hole,
an extremal Reissner-Nordstr\"om black hole will appear with $(M-\sum_{i}\omega_i)^2-Q^2=0$,
where $Q$ denotes its charge.
In this case, further thermal and Hawking radiation emission of particles
will stop when the temperature of an extremal black hole reaches $T=0$.
This final state of a Reissner-Nordstr\"om black hole is the extremal black hole with
mass $m=M-\sum_{i}\omega_i$ and charge $Q$.
The number of the microstates for the Reissner-Nordstr\"om black hole is therefore given by
\begin{eqnarray}\label{rn:to:extremal}
\Omega_{\rm initial}\left(M,Q\right)=\left[\prod_{i=1}^{n}\Omega(\omega_i)\right]\cdot\Omega_{\rm final}(m,Q),
\end{eqnarray}
where $\Omega_{\rm final}(m,Q)$ denotes the number of microstate for
a charged black hole with mass $m$ and charge $Q=m$. Quite generally, such
an extremal black hole can be viewed as the ground state for a charged black hole with mass $m$ and charge $Q=m$.

Since an extremal black hole no longer emits particles,
we cannot find its number of microstates by counting the corresponding microstates of its radiations.
In order to obtain the number of microstates for a Reissner-Nordstr\"om black hole self-consistently,
we will consider charged particle emissions which can carry away both mass and charge.
Analogously, the number of microstates for a Reissner-Nordstr\"om black hole can be obtained as
\begin{eqnarray}\notag 
\Omega_{\rm initial}(M, Q)=\prod_{i=1}^{n}\Omega_{\rm radiation}(\omega_i,q_i)
=\prod_{i=1}^{n}\frac{1}{\Gamma(\omega_i,q_i)},
\end{eqnarray}
where $\Gamma(\omega,q)$ is the radiation probability for the tunneling of a
particle with mass $\omega$ and charge $q$~\cite{zz05},
\begin{eqnarray} 
\Gamma(\omega,q)&=&\exp\Big[\pi\big(M-\omega+\sqrt{(M-\omega)^2-(Q-q)^2}\,\big)^2 \notag \\
&&-\pi\big(M+\sqrt{M^2-Q^2}\,\big)^2\Big].
\end{eqnarray}
It is easy to verify that this tunneling rate spectrum satisfies the iff condition
Eq.~(\ref{boundary:condition}), and thus can be utilized to calculate the number
of microstates of a Reissner-Nordstr\"om black hole. After straightforward calculations,
we find
\begin{eqnarray}\label{number:rn}
\Omega_{\rm initial}(M, Q)=\prod_{i=1}^{n}\frac{1}{\Gamma(\omega_i,q_i)}
=e^{\pi(M+\sqrt{M^2-Q^2}\,)^2}.
\end{eqnarray}
With this result for the number of microstates, we obtain the entropy
for a Reissner-Nordstr\"om black hole as
\begin{displaymath}
S=\ln \Omega_{\rm initial}(M, Q)=\pi(M+\sqrt{M^2-Q^2}\,)^2.
\end{displaymath}
Again this confirms that the Bekenstein-Hawking entropy for the Reissner-Nordstr\"om black hole
can be obtained by counting the number of microstates of charged Hawking radiations.

\section{extremal Reissner-Nordstr\"om black hole}

Now for the simplest case of an extremal black hole, i.e., the extremal Reissner-Nordstr\"om black
hole, which is at zero temperature but with a nonzero area horizon, it is interesting to find out
whether such an extremal black hole has an {\it area} entropy or not.
Some interesting results on this were obtained from string theory~\cite{as95}, where
Bekenstein-Hawking entropies for several specific extremal black holes are calculated
through explicit counting of the degeneracies of BPS soliton bound states~\cite{sv96,sl05}.
In the following, these results are recovered by us using the semiclassical emission spectrum,
which satisfies the iff condition Eq. (\ref{boundary:condition}),
based on the simple approach along the lines developed in the above.

As described in the previous section an extremal Reissner-Nordstr\"om black hole can result from
a Reissner-Nordstr\"om black hole if only neutral particles are emitted. Equation~(\ref{rn:to:extremal})
shows that the number of microstates for an extremal black hole $\Omega_{\rm final}(m,Q)$ can be
determined if the number of microstates of the initial black hole $\Omega_{\rm initial}(M,Q)$ and
the number of microstates of all Hawking radiations $\prod_{i=1}\limits ^{n}\Omega(\omega_i)$ are
known, i.e.,
\begin{displaymath}
\Omega_{\rm final}(m,Q)=\frac{\Omega_{\rm initial}(M,Q)}{\prod\limits_{i=1}^{n}\Omega(\omega_i)}.
\end{displaymath}
The number of microstates $\Omega_{\rm initial}(M,Q)$ is already
obtained from the microstates of charged emissions in the previous section, as shown in Eq. (\ref{number:rn}).
The number of microstates for neutral Hawking radiations is inverse of the the semi-classical
tunneling rate, which can be calculated based on Hawking radiation as tunneling~\cite{pw00}
\begin{displaymath}
\Gamma(\omega)=e^{-2\pi[2\omega(M-\omega/2)-(M-\omega)\sqrt{(M-\omega)^2-Q^2}+M\sqrt{M^2-Q^2}]}.
\end{displaymath}
It is easy to verify that this radiation spectrum satisfies the iff condition Eq.~(\ref{boundary:condition}).
A straightforward calculation shows that after a Reissner-Nordstr\"om black hole with mass $M$ and charge $Q$
is reduced to an extremal black hole with $m=Q$, the number of microstates for its emissions is
\begin{eqnarray}
\prod_{i=1}^{n}\Omega(\omega_i)=e^{2\pi(M^2-m^2-m\sqrt{m^2-Q^2}+M\sqrt{M^2-Q^2})},\notag
\end{eqnarray}
The number of microstates for the remnant extremal black hole is thus found to be
\begin{eqnarray}
\Omega_{\rm final}(m,Q)=\frac{\Omega_{\rm initial}(M,Q)}{\prod\limits_{i=1}^{n}\Omega(\omega_i)}
=e^{\pi(m+\sqrt{m^2-Q^2})^2},\notag
\end{eqnarray}
which further simplifies to $\Omega_{\rm final}(m,Q=m)=e^{\pi m^2}$ for an extremal Reissner-Nordstr\"om black hole.
Putting these all together, the entropy for an extremal Reissner-Nordstr\"om black hole
with mass $m$ and charge $Q=m$ can be obtained as
\begin{eqnarray}
S=\ln \Omega_{\rm final}(m,Q=m)=\pi m^2,
\end{eqnarray}
which is exactly equal to result obtained from string theory~\cite{sl05}.

\section{other black holes}

For a Kerr black hole, its Hawking radiation particles can carry off both energy and angular momentum \cite{jwc06}.
In the end when the black hole itself is exhausted, we can count the number of microstates of its Hawking
radiations, which according entropy conservation for unitary evolution, should rightfully be the same
as the number of microstates for the initial Kerr black hole. For a Kerr-Newman black hole, we should consider
its emissions carrying energy, charge, and angular momentum, in order to avoid the appearance
of an extremal black hole at the final stage of Hawking radiation~\cite{vad11}.
Its number of microstates can also be obtained by counting the
number of microstates of the emitted Hawking radiation particles.
As long as unitary is maintained for Hawking radiation process,
 in principle, the entropies for all non-extremal black holes can be computed along the same lines
illustrated in previous sections
 by counting the numbers of microstates for their Hawking radiations.

Extremal black holes do not emit particles, thus their microstates cannot be simply
counted from their Hawking radiations. But an extremal black hole can be viewed as resulting
from its corresponding non-extremal black hole through Hawking radiation.
For instance, an extremal Reissner-Nordstr\"m black hole is the final state of its corresponding
Reissner-Nordstr\"m black hole. The number of microstates for an extremal black hole
thus can be computed as the quotient of the number of microstates for the initial black hole
and the number of microstates for all neutral Hawking radiations emitted by the
initial black hole before ending up as the extremal black hole.
Thus, at least in principle, using the semiclassical approach establishes
in previous sections we can obtain the number of microstates and the entropies for
all other extremal black holes.

\section{Discussion and conclusion}

We establish a necessary and sufficient (iff) condition for falsifying
unitary Hawking radiation through a simple check of two
emission probabilities governed by the corresponding Hawing radiation spectrum.
This condition can be easily used to rule out quantum gravity theories.
We show that the numbers of microstates for non-extremal black holes can be computed through counting the
microstates of their Hawking radiations.
As illustrative examples we show the precise values of Bekenstein-Hawking
entropies for Schwarzschild black holes and Reissner-Nordstr\"om black holes can
be calculated from the counted numbers of microstates.
Of particular interest, the number of microstate and the corresponding entropy we
find for the extremal Reissner-Nordstr\"om black hole, are consistent with the string theory results \cite{sv96,sl05,as95}.
Since Hawking radiation occurs near event horizon, our result implies black hole \emph{area}
entropy is only related to the microstates near event horizon, and all information about the collapsed
matter in the hole can be carried off with Hawking radiations near event horizon.

It has been questioned whether extremal black holes possess entropies or not since they are at
zero temperature yet with nonzero horizon areas~\cite{hhr95,gm95}.
We obtain the number of microstates for an example of extremal black hole by counting the number
of the microstates of the alternative Hawking radiations for its corresponding (initial non-extremal) black hole.
The nontrivial number of microstates we obtain as well as the nontrivial entropy for the
extremal black hole illustrated, or an extremal Reissner-Nordstr\"om black hole, implies
there exists interior structure for the extremal black hole with degenerate energy levels.
This study thus shines new light on a resolution to the dispute
over the existence of entropies for extremal black holes in the semiclassical limit.

\section*{ACKNOWLEDGEMENT}

This work is supported by MOST 2013CB922003 and 2013CB922004 of
the National Key Basic Research Program of China,
and by the NSFC (No.~91121005, No.~91421305, No.~11374176, No.~11328404,
and No.~61471356).

\end{document}